\documentclass{wlscirep}
\usepackage[utf8]{inputenc}
\usepackage[T1]{fontenc}
\usepackage{bm} 
\usepackage{placeins}
\usepackage{xcolor}
\usepackage{comment}
\usepackage{subcaption}

\title{QUBO-Based Optimization of Social Indicator Configurations for Working-Age Population Growth}

\author[1]{Hayate Wada}
\author[2]{Seiya Miyamoto}
\author[2]{Tatsumasa Ogawa}
\author[2]{Kazuki Uehara}
\author[2,*]{Masaru Hitomi}
\author[2, 3, 4, 5]{Masayuki Ohzeki}
\affil[1]{Graduate School of Engineering, Tohoku University, Sendai 980-8579, Japan}
\affil[2]{Graduate School of Information Sciences, Tohoku University, Sendai 980-8579, Japan}
\affil[3]{Department of Physics, Institute of Science Tokyo, Tokyo 152-8551, Japan}
\affil[4]{Research and Education Institute for Semiconductors and Informatics, Kumamoto University, Kumamoto 860-8555, Japan}
\affil[5]{Sigma-i Co., Ltd., Tokyo 108-0075, Japan}

\affil[*]{masaru.hitomi.e5@tohoku.ac.jp}

\begin{abstract}
The decline of the working-age population is a major challenge for regional sustainability, particularly in ageing societies such as Japan.
Although municipal demographic change is associated with multiple socioeconomic indicators, 
it remains difficult to translate such associations into interpretable and optimization-ready models.
We present the first methodological demonstration of a quadratic unconstrained binary optimization (QUBO)-based framework for exploring social-indicator configurations associated with working-age population growth.
Using Japanese municipal data, we regressed the 2010--2020 working-age population growth rate on ten social indicators encoded as discretized one-hot variables.
The resulting quadratic surrogate model achieved reasonable predictive performance, with a test-set correlation coefficient of 0.84 and an average coefficient of determination of $R^{2}=0.76$.
Because the model is quadratic, its coefficients can be represented as an interpretable matrix whose diagonal elements describe individual indicator-level contributions and whose off-diagonal elements describe pairwise associations between indicator levels.
We then converted the fitted surrogate model into a QUBO formulation with one-hot constraints and optimized it using quantum annealing, simulated annealing, and Gurobi.
All three approaches identified the same optimal feasible configuration in the present problem setting, while the annealing-based samplers also generated feasible suboptimal configurations with different predicted growth rates.
Municipality-level single-indicator flip analyses further showed that modifying one indicator can either increase or decrease the predicted growth rate depending on the configuration of other indicators.
These results demonstrate that QUBO-based modelling can provide an interpretable and optimization-ready framework for connecting municipal social statistics, nonlinear indicator interactions, and model-based scenario generation.
The proposed framework should be interpreted as an exploratory tool for policy discussion rather than as a causal estimate of policy interventions.

\end{abstract}

\begin{document}

\flushbottom
\maketitle

\thispagestyle{empty}


\section*{Introduction}

Many advanced economies are entering a phase in which demographic change directly constrains economic potential~\cite{OECD2025EmploymentOutlook}.
A central mechanism is the contraction of the working-age population, which is projected to intensify over coming decades in many OECD countries~\cite{OECD2025EmploymentOutlook}.
This shift is not spatially uniform: population decline and ageing often concentrate in peripheral, rural, and small-to-medium urban areas, where out-migration and compositional ageing amplify local labour and skill shortages and erode fiscal capacity~\cite{OECD2025ShrinkingSmartly}.
These spatial disparities are increasingly recognised as a major policy challenge because they interact with service provision, infrastructure maintenance, and local economic resilience~\cite{OECD2025ShrinkingSmartly}.

Japan is an instructive setting for studying this problem quantitatively.
National demographic headwinds coincide with pronounced subnational divergence, including continued concentration in the Tokyo metropolitan area and accelerated decline in many non-metropolitan municipalities~\cite{IPSS2018RegionalProjections,StatisticsBureauJapan2020Census}.
Official municipal population projections provide a consistent reference for the magnitude and geography of future change, while recent empirical work highlights that ``shrinkage'' dynamics are shaped by heterogeneous local conditions rather than by a single demographic driver~\cite{IPSS2018RegionalProjections,Kato2023ShrinkingCities}.

Despite broad agreement that declining working-age population growth is a critical issue, turning this diagnosis into actionable, locality-specific guidance remains difficult.
First, local demographic outcomes are influenced by multiple social and economic dimensions, such as employment structure, accessibility of services, education, and family-related conditions, that may interact nonlinearly.
Second, policymaking typically operates under combinatorial constraints: improving one indicator may require complementary changes in other indicators to produce measurable demographic effects.
Empirical studies have therefore moved beyond simple bivariate associations, using probabilistic or machine-learning frameworks to identify the relative importance of multiple indicators for population change~\cite{Kato2023ShrinkingCities}.

However, two gaps persist in the current literature.
The first is interpretability in a form that can directly support policy design: while flexible models can predict or rank correlates, it is often nontrivial to translate their outputs into a transparent objective function that suggests which combination of indicator levels should be targeted.
The second is the optimization step itself.
Even when a model implies a desirable direction for each variable, the best joint configuration is a discrete, high-dimensional search problem when indicators are binned or policy-relevant levels are considered.
Although exact optimization is possible for moderate problem sizes, the number of feasible configurations grows exponentially with the number of indicators and discretization levels.
This motivates an optimization framework that can explore a large discrete landscape and return not only a single optimum but also feasible candidate configurations.

Here we address these gaps by constructing an interpretable optimization pipeline that links municipal working-age population growth to multiple social indicators and then searches for indicator configurations associated with higher predicted growth.
Our approach is built around a quadratic unconstrained binary optimization (QUBO) representation, which provides a compact and explicit objective function over binary decision variables.
QUBO models are broadly used to formulate combinatorial optimization problems and can be related to Ising Hamiltonians~\cite{Kochenberger2014UBQP,Glover2019QUBO,Lucas2014Ising}.
They can also be handled by exact solvers, classical annealing heuristics, and quantum annealing methods~\cite{Kirkpatrick1983SA,Kadowaki1998QA,Johnson2011QuantumAnnealing,Gurobi2026}.
This property makes QUBO a useful representation for comparing different optimization approaches while retaining an interpretable objective function.
Applications such as traffic-signal optimization illustrate how Ising/QUBO formulations can connect real-world objectives to annealing-based solvers~\cite{Inoue2021TrafficQA}.

Using a dataset of Japanese municipalities, we construct a quadratic surrogate model $G(\vec{x})$ to describe the relationship between the working-age population growth rate and social indicators.
In this study, social indicators from the 2010 Population Census are used as explanatory variables, and the working-age population growth rate between 2010 and 2020 is used as the target variable~\cite{StatisticsBureauJapan2010Census,StatisticsBureauJapan2020Census}.
Because the model is formulated as a quadratic function, it captures not only the individual contribution of each indicator level but also pairwise interactions between indicator levels.
The learned coefficients can be visualized as a coefficient matrix, allowing us to inspect both individual indicator-level contributions and pairwise compatibility between social conditions.

We then optimize the QUBO representation of the surrogate model using simulated annealing, quantum annealing, and Gurobi~\cite{Kirkpatrick1983SA,Kadowaki1998QA,Gurobi2026}.
Gurobi is used as an exact reference optimizer for the present problem size, while the annealing-based methods are used to sample feasible configurations in the QUBO landscape.
This comparison is not intended to claim a computational advantage of quantum annealing for the present small-scale problem.
Rather, it is used to validate the consistency of the QUBO formulation and to examine whether annealing-based searches can recover the optimal solution and provide additional feasible candidate configurations.
Such candidate configurations are useful because the globally optimal state may not be immediately achievable in practical municipal planning.

Finally, to connect the model output to local interpretation, we perform municipality-level single-indicator flip analyses.
For selected municipalities, we evaluate how the predicted growth rate changes when one social-indicator level is modified while all other indicators are kept fixed.
These simulations should be interpreted as model-based local sensitivity analyses, rather than causal estimates of policy interventions.
They illustrate how an optimization-ready and interpretable surrogate function can be used not only to analyse correlates retrospectively but also to generate candidate scenarios for localized policy discussion.

We regard this study as a methodological demonstration rather than a causal evaluation of demographic policy.
The aim is to show that associations learned from municipal social statistics can be encoded as an interpretable quadratic objective and explored using QUBO-based optimization.
In doing so, our study contributes a methodology that integrates three components: 
subnational demographic-policy relevance, multifactor quantitative modelling of working-age population growth, and QUBO-based combinatorial optimization over discretized social-indicator configurations.
This framework provides a transparent way to connect municipal social statistics, interpretable quadratic modelling, and optimization-based exploration of feasible indicator configurations.

\section*{Methods}

\subsection*{QUBO formed surrogate model and data Preprocessing}

We employed a quadratic function $G(\vec{x})$ to define the growth rate of the working-age population growth rate from social indicators as followes;

\begin{equation} \label{eq:G}
    G(\vec{x}) = \sum_{(i,n)<(j,m)}  A_{ijnm} x_{in} x_{jm} + \sum_{i,n}A_{in}x_{in} +A_0,
\end{equation}
where $x_{in} \in \{0,1\}$ is a binary variable indicating whether the $n$-th level of the $i$-th social indicator is selected, $A_{ijnm} \in \mathbb{R}$ represents the interaction strength between $x_{in}$ and $x_{jm}$, $A_{in}$  denotes the linear coefficient (local bias) associated with $x_{in}$, $A_0$ is a constant offset (global bias) independent of the variables.

Each data point can be represented as a vector of social indicators. However, the components of these indicators are generally continuous-valued. Therefore, in order to apply Eq.~\eqref{eq:G}, it is necessary to convert them into binary variables. To this end, we adopt a two-step procedure consisting of discretization and encoding.
First, each continuous-valued indicator $i$ is transformed into a discrete variable. 
Specifically, the entire dataset is sorted in ascending order, and the values of indicator $i$ are divided into four levels based on quartiles, assigning integers from 1 to 4. 
We use four levels because quartile-based discretization gives approximately balanced groups for each indicator, preserves a simple ordinal interpretation, and keeps the number of binary variables manageable for the present QUBO demonstration.
Next, these discrete levels are encoded using four binary variables. Concretely, $x_{in}$ takes the value 1 if the $i$-th indicator belongs to level $n$, and 0 otherwise. 
In this way, indicator $i$ can be represented as a one-hot vector $\hat{x}_i=(x_{i1},x_{i2},x_{i3},x_{i4})^\mathsf{T}$. 
For example, if the $i$-th indicator is assigned to level 2, it is represented as $\hat{x}_i = (0,1,0,0)^\mathsf{T}$. 
Through this process, the set of $F$ social indicators associated with each municipality can be represented as a binary vector $\vec{x} = (\hat{x}_1, \ldots, \hat{x}_F)^\mathsf{T} \in \{0,1\}^{4F}$.




\subsection*{Regression model}
Each dataset consists of a preprocessed social indicator vector $\vec{x}$, as described in the previous section, and the corresponding working-age population growth rate $y$.
Our objective is to perform regression on a given dataset $\mathcal{D} = \{ (\vec{x}^{(d)}, y^{(d)}) \}_{d=1}^{D}$ in order to determine the parameters $A_{ijnm}$ and $A_0$, and thereby construct the surrogate function defined in Eq.~\eqref{eq:G}.
To this end, we adopt a framework based on Bayes' theorem:

\begin{equation}
P(\vec{a} \mid \vec{x},y) = \frac{P(y \mid \vec{a},\vec{x})\, P(\vec{a} )}{P(y \mid \vec{x})},
\end{equation}
where $\vec{a} = (A_0, A_{11}, \ldots, A_{F4}, A_{1112}, \ldots, A_{FF34}) \in \mathbb{R}^{P}$ denotes the vector of model parameters, with $P = 1 + 4F + \binom{4F}{2}$ being the total number of parameters. 
Here, $P(\vec{a} \mid \vec{x}, y)$ is the posterior distribution, $P(y \mid \vec{a}, \vec{x})$ is the likelihood function, $P(\vec{a})$ is the prior distribution, and $P(y \mid \vec{x})$ is the marginal likelihood.






In this study, we estimate the parameters by maximizing the posterior distribution, i.e., using maximum a posteriori (MAP) estimation. 
Taking the logarithm of Eq.~(2), we obtain
\begin{align}
\vec{a}^* 
    &= \arg\max_{\vec{a}} \, P(\vec{a} \mid \vec{x}, y) \notag \\
    &= \arg\max_{\vec{a}} \, P(y \mid \vec{a}, \vec{x}) \, P(\vec{a}) \notag \\
    &= \arg\max_{\vec{a}} \left\{ \log P(y \mid \vec{a}, \vec{x}) + \log P(\vec{a}) \right\}.
    \label{eq:a*}
\end{align}
Let us introduce a feature vector $X = (1, x_{11}, \ldots, x_{F4}, x_{11}x_{12}, \ldots, x_{F3}x_{F4})^{\top} \in \{0,1\}^{P}$. Then, from Eq.~\eqref{eq:G}, the surrogate function can be written as $G(\vec{x}) = \vec{a}^{\top} X$. For each data $\vec{x}^{(d)}$, we construct the corresponding feature vector $X^{(d)}$. Assuming that the error between the true value $y^{(d)}$ and the model prediction $G(\vec{x}^{(d)}) = \vec{a}^{\top} X^{(d)}$ follows a standard normal distribution, the likelihood in Eq.~(2), $P(y \mid \vec{a}, \vec{x})$, can be expressed as
\begin{align}
    P(y \mid \vec{a}, \vec{x}) 
    &= \prod_{d=1}^{D} \frac{1}{\sqrt{2\pi}} 
    \exp\left( -\frac{1}{2} \left( y^{(d)} - \vec{a}^{\top} X^{(d)} \right)^2 \right) \notag \\
    &= \left( \frac{1}{\sqrt{2\pi}} \right)^{D}
    \exp\left( -\frac{1}{2} \sum_{d=1}^{D} \left( y^{(d)} - \vec{a}^{\top} X^{(d)} \right)^2 \right).
    \label{eq:likelihood_function}
\end{align}
On the other hand, we assume that the prior probability distribution $P(a)$ also follows a standard normal distribution, 
\begin{equation}
    P(\vec{a}) = \sqrt{\frac{\lambda}{2\pi}} \exp \left( -\frac{\lambda}{2} \vec{a}^{\top} \vec{a} \right),
    \label{eq:prior_distribution}
\end{equation}
where $\lambda$ denotes a positive regularization parameter that controls the variance of the prior distribution. This parameter determines the strength of the penalty imposed on the magnitude of the parameter vector $\vec{a}$, thereby reflecting prior assumptions on model complexity.
Finally, by combining Eqs.~\eqref{eq:likelihood_function} and ~\eqref{eq:prior_distribution}, Eq.~\eqref{eq:a*} can be rewritten as follows:
\begin{align}
    \vec{a}^* &= \arg \max_{\vec{a}} \left( -\frac{1}{2} \sum_{d=1}^{D} \left( y^{(d)} - \vec{a}^{\top} X^{(d)} \right)^2 - \frac{\lambda}{2} \vec{a}^{\top} \vec{a} \right) \notag\\
    &= \arg \min_{\vec{a}} \left( \frac{1}{2} \sum_{d=1}^{D} \left( y^{(d)} - \vec{a}^{\top} X^{(d)} \right)^2 + \frac{\lambda}{2} \vec{a}^{\top} \vec{a} \right) \label{eq:regression_parameter}.
\end{align}
This formulation shows that maximizing the posterior probability is equivalent to solving a regularized least-squares problem with an $L_2$ penalty.
We note that, in the present proof-of-concept analysis, we have adopted the $L_2$ penalty as the simplest isotropic regularization assumption on the regression coefficients.

\subsection*{Annealing-based optimization}
Since the surrogate function obtained through regression is quadratic in form, 
it can be transformed into a QUBO cost function whose minimization corresponds to maximizing $G(\vec{x})$ under the one-hot constraints.
The cost function in the Quadratic Unconstrained Binary Optimization (QUBO) formulation to be minimized by annealing-based methods and an exact reference solver is defined as follows;
\begin{align}
H(\vec{x}) &= - G(\vec{x}) + \alpha \sum_{i=1}^{F} \left( \sum_{n=1}^{4} x_{in} - 1 \right)^2 \nonumber \\
&= - \left( \sum_{(i,n)<(j,m)}  A_{ijnm} x_{in} x_{jm} + \sum_{i,n}A_{in}x_{in} +A_0 \right) + \alpha \sum_{i=1}^{F} \left( \sum_{n=1}^{4} x_{in} - 1 \right)^2
\label{eq:H}
\end{align}
where the first term represents the objective function defined in Eq.~\eqref{eq:G}. The second term is a factor that imposes a one-hot constraint on $x$ and $\alpha$ is a hyperparameter that controls the strength of this constraint. This one-hot constraint arises from the discretization and binarization of the variables $x$, and serves to prevent multiple levels from being simultaneously active for a single municipality.
The combination of variables that minimizes $H$ corresponds to the configuration that maximizes the surrogate growth-rate function $G(\vec{x})$ while satisfying the one-hot constraints.
Therefore, by exploring the states that minimize $H$ using simulated and quantum annealing, it becomes possible to identify the combination of social indicators that maximizes the working-age population growth rate.

\section*{Results and Discussion}

\subsection*{Dataset}
Ten social indicators of Japanese municipalities were used as explanatory variables $\vec{x}$, obtained from the official government statistics portal e-Stat~\cite{eStatPortal,eStat2010Census}.
These indicators are summarized in Table~\ref{tab:1}, and are based on the 2010 Population Census.

The target variable $y$ was defined as the growth rate of the working-age population (aged 15–64) between 2010 and 2020. 
The analysis included 1,882 municipalities across Japan, excluding those that merged or dissolved during this period and Futaba Town in Fukushima Prefecture, for which data were unavailable due to the Great East Japan Earthquake.  

To improve regression accuracy, 
outliers were removed using the interquartile range (IQR) criterion, excluding municipalities whose growth rates fell outside the range of $[Q_1 - 1.5 \times IQR, Q_3 + 1.5 \times IQR]$. 
After this filtering, 1,855 municipalities remained for regression analysis.
This filtering was intended to remove municipalities whose growth rates were likely dominated by exceptional exogenous shocks or strong policy-driven population movements, rather than by the ordinary socioeconomic indicators considered in this study.

\begin{table}[h]
\centering
\caption{List of social indicators used in the analysis. The numbers in the table correspond to the subscript $n$ of $\hat{x}_n$.
}
\label{tab:1}
\begin{tabular}{clll}
\hline
No. & Label & Social Indicator & Unit \\
\hline
1 & Population & Population density & (persons/km$^2$) \\
2 & Age & Average age & (years) \\
3 & Sex & Sex ratio & (males per 100 females) \\
4 & Foreign & Ratio of foreign residents & (\%) \\
5 & Single & Ratio of single-person households & (\%) \\
6 & Three & Ratio of three-generation households & (\%) \\
7 & Primary & Ratio of primary industry workers & (\%) \\
8 & Secondary & Ratio of secondary industry workers & (\%) \\
9 & Tertiary & Ratio of tertiary industry workers & (\%) \\
10 & Day-night & Daytime-nighttime population ratio & -- \\
\hline
\end{tabular}
\end{table}

\newpage

\subsection*{Performance and interpretability of the surrogate model}

\begin{figure}[t]
    \centering
    \includegraphics[width=0.95\linewidth]{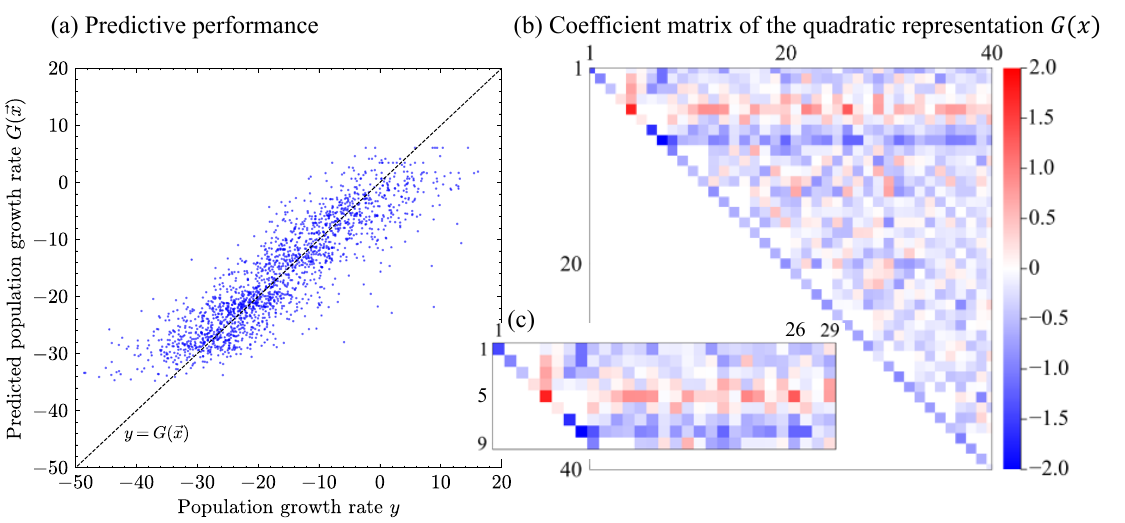}
    \caption{
    Performance and coefficient structure of the quadratic surrogate model $G(\vec{x})$.
    (a) Observed working-age population growth rate $y$ versus the fitted value $G(\vec{x})$ for all municipalities.
    The dashed line represents the ideal relation $y=G(\vec{x})$.
    The test-set correlation coefficient was 0.84, with an average mean squared error of 33.36 and an average coefficient of determination of $R^{2}=0.76$ over 20 train--test splits.
    (b) Heatmap of the coefficient matrix of $G(\vec{x})$ for the 40 binary variables obtained from the one-hot encoding of ten social indicators.
    Diagonal elements correspond to individual binary-variable contributions, and off-diagonal elements correspond to pairwise interactions.
    Positive coefficients increase the predicted growth rate $G(\vec{x})$.
    (c) Enlarged view of the selected region.
    The fifth diagonal element, $Q_{5,5}=1.75$, corresponding to the lowest average-age quartile, gives the largest diagonal contribution.
    The element $Q_{5,26}=1.30$, corresponding to the interaction between the lowest average-age quartile and the second quartile of the primary-industry worker ratio, gives the largest off-diagonal contribution in the enlarged region.
    }
    \label{fig:QUBO}
\end{figure}

We first examined whether the quadratic surrogate model $G(\vec{x})$ provides a sufficiently accurate representation of the working-age population growth rate.
After the preprocessing described above, the dataset was divided into training and test sets with a ratio of 19:1.
This train--test split was repeated 20 times for cross-validation.
During this process, the regularization hyperparameter $\lambda$ was determined by evaluating several candidate values, and the value that minimized the average mean squared error (MSE) across the test sets was selected.
For each split, the model parameters were estimated from the training set, and the prediction accuracy was evaluated on the corresponding test set.
Figure~\ref{fig:QUBO}(a) shows the fitted values of $G(\vec{x})$ for all municipalities, in order to visualize the overall agreement between the observed and estimated growth rates.
The quantitative predictive performance was evaluated using the test sets.
The average mean squared error over the 20 test sets was 33.36, and the average coefficient of determination was $R^{2}=0.76$.
The correlation coefficient between the observed and predicted values on the test sets was 0.84.
These results indicate that the fitted quadratic function captures the overall trend of the working-age population growth rate reasonably well.
The purpose of this model is not to provide a complete demographic forecast, but to construct an interpretable surrogate objective function that can be used for subsequent combinatorial optimization.
To interpret the fitted surrogate model, we visualized the coefficients of $G(\vec{x})$ as a matrix.
Since ten social indicators were encoded into four binary variables each, the model contains 40 binary variables in total.
Let $x_k$ denote the flattened binary variable corresponding to $x_{in}$, where $k=4(i-1)+n$.

Figure~\ref{fig:QUBO}(b) shows the upper-triangular part of this coefficient matrix $Q$.
It should be emphasized that this matrix represents the coefficients of $G(\vec{x})$, not those of $-G(\vec{x})$.
Therefore, a larger positive coefficient in Fig.~1(b) contributes to increasing the predicted working-age population growth rate.
In contrast, when the model is converted into the QUBO cost function $H(\vec{x})=-G(\vec{x})+\cdots$, the sign of the contribution to the minimized energy is reversed.
The diagonal elements $Q_{kk}$ represent the individual contributions of the corresponding binary variables.
Among the diagonal elements, the largest value was found at $Q_{5,5}=1.75$.
The fifth variable $x_5$ corresponds to the first quartile of the average age indicator.
In other words, $x_5=1$ indicates that the municipality belongs to the lowest average-age group.
This positive diagonal coefficient suggests that municipalities with lower average age are strongly associated with higher predicted working-age population growth in the fitted surrogate model.
This tendency is qualitatively consistent with the fact that the target variable is the growth rate of the working-age population, because municipalities with younger demographic structures are expected to have a greater capacity to maintain or increase their working-age population.
The off-diagonal elements $Q_{kl}$ represent pairwise interaction effects between two binary variables.
A positive off-diagonal coefficient indicates that the simultaneous activation of the corresponding two variables increases $G(\vec{x})$ beyond their individual diagonal contributions.
Conversely, a negative off-diagonal coefficient indicates that the corresponding combination is associated with a lower predicted growth rate.
As shown in Fig.~\ref{fig:QUBO}(b), relatively large off-diagonal coefficients tend to appear around variables whose diagonal elements have large absolute values.
This suggests that variables with strong individual contributions also participate in important pairwise interactions.

Figure~\ref{fig:QUBO}(c) shows an enlarged view of the region including the average-age variables and their interactions with other indicators.
In this region, the coefficient at row 5 and column 26 has the largest off-diagonal value, $Q_{5,26}=1.30$.
This element corresponds to the interaction between the lowest average-age quartile and the second quartile of the primary-industry worker ratio.
The large positive value of $Q_{5,26}$ suggests that the combination of a low average age and a moderate primary-industry worker ratio is strongly associated with an increase in the predicted working-age population growth rate.
This result illustrates the role of the quadratic terms in the surrogate model: they capture not only the independent effect of each social indicator level, but also the compatibility between different indicator levels.
The coefficient matrix should be interpreted as an association structure learned by the fitted surrogate model, rather than as a set of causal effects.
In particular, changing a single social indicator in an actual municipality may involve social, economic, and institutional constraints that are not explicitly included in the present model.
Moreover, because each social indicator is represented by a one-hot vector, different levels of the same indicator cannot be simultaneously active in feasible configurations.
Therefore, the most direct interpretation of the off-diagonal elements is obtained for interactions between different social indicators.

Overall, Fig.~\ref{fig:QUBO} demonstrates that the surrogate model has both predictive validity and interpretability.
The fitted coefficient matrix provides an explicit landscape over binary indicator configurations, which serves as the objective function to be optimized by annealing-based methods in the next section.

\subsection*{Optimization of the QUBO model by annealing methods}

We next optimized the binary indicator configuration using the QUBO formulation described in the Methods section.
The penalty coefficient for the one-hot constraint was set to $\alpha = 3$ based on preliminary experiments. 
At this value, all samples generated by simulated annealing satisfied the one-hot constraint, indicating that the penalty coefficient was sufficient to enforce the constraint. Increasing the penalty coefficient further would cause the penalty term to dominate the objective function, potentially degrading the search performance with respect to the original optimization objective.
In this formulation, 
feasible configurations satisfying the one-hot constraint are evaluated by the surrogate function $G(\vec{x})$.
Therefore, although the annealing procedures minimize the corresponding QUBO cost function, the results are shown in terms of $G(\vec{x})$, 
which directly represents the predicted working-age population growth rate.
We compared three optimization approaches: quantum annealing, simulated annealing, and Gurobi.
Quantum annealing (QA) was performed using the D-Wave Advantage2 system, while simulated annealing (SA) was implemented using the Neal SimulatedAnnealingSampler. For both methods, the default parameter settings were used except that the number of reads was set to 1,000. Accordingly, the default annealing time of $20 \; \mu s$ was used for QA, whereas the default annealing schedule of 1,000 Monte Carlo sweeps was used for SA.
Gurobi was used as an exact reference optimizer for the present QUBO problem.
Only samples satisfying the one-hot constraint were included in the histogram shown in Fig.~2.
All samples obtained by SA were feasible solutions, whereas $39.4 \;\%$ of the samples obtained by QA were infeasible.
This is considered to be primarily due to the need to embed the QUBO onto the sparse Zephyr graph architecture of the D-Wave Advantage2 system, where each logical qubit is represented by a chain of multiple physical qubits. The default chain strength was used throughout this study. Preliminary experiments showed that adjusting the chain strength did not significantly improve the fraction of feasible solutions.

\begin{figure}[h]
    \centering
    \includegraphics[width=0.8\linewidth]{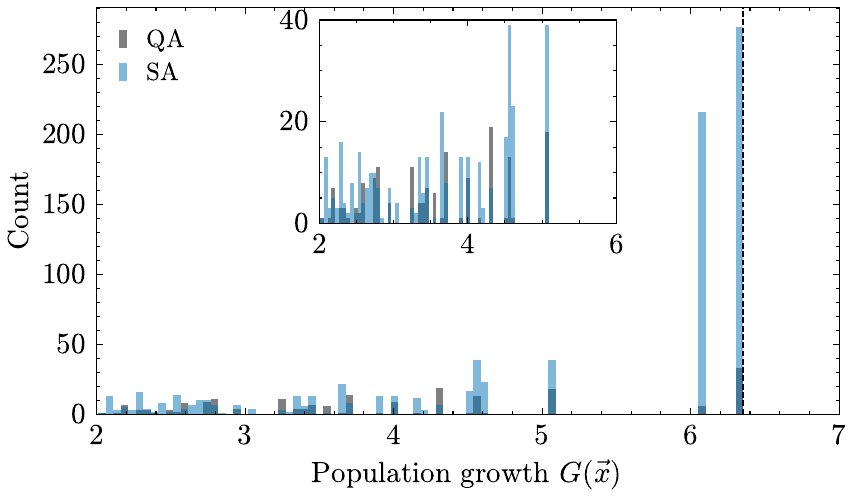}
    \caption{
        Sampling distributions obtained by simulated annealing and quantum annealing. 
        The horizontal axis represents the surrogate value $G(\vec{x})$, corresponding to the predicted working-age population growth rate, and the vertical axis represents the number of samples.
        Only configurations satisfying the one-hot constraints are included.
        The vertical dashed line indicates the exact optimum obtained by Gurobi.
        The inset enlarges the low-count region to show feasible suboptimal configurations sampled by the annealing methods.
        }
    \label{fig:histgram}
\end{figure}

Figure~\ref{fig:histgram} shows the distributions of feasible samples obtained by quantum annealing and simulated annealing.
The horizontal axis represents the value of $G(\vec{x})$, and the vertical axis represents the number of samples.
The vertical dashed line indicates the exact optimum obtained by Gurobi.
Both quantum annealing and simulated annealing reached the same optimal value as Gurobi, $G_{\mathrm{opt}}=6.354 \cdots$.
This agreement indicates that the QUBO formulation consistently encodes the maximization problem of the surrogate function and that the annealing-based searches can identify the optimal feasible configuration in the present setting.
The optimal indicator configuration was
\begin{align}
[4,1,4,4,4,2,1,1,4,4] .
\end{align}
Thus the optimal configuration consists of the fourth level of population density, 
the first level of average age, 
the fourth level of sex ratio, 
the fourth level of the ratio of foreign residents, 
the fourth level of the ratio of single-person households, 
the second level of the ratio of three-generation households, 
the first level of the ratio of primary-industry workers, 
the first level of the ratio of secondary-industry workers, 
the fourth level of the ratio of tertiary-industry workers, 
and the fourth level of the daytime--nighttime population ratio.
This configuration should be interpreted as the state that maximizes the fitted surrogate function $G(\vec{x})$ under the discretized one-hot representation, rather than as a direct causal prescription for municipalities.

The inset of Fig.~\ref{fig:histgram} enlarges the region where the number of samples is small.
This enlarged view is useful because annealing methods do not only provide the single best configuration but also sample multiple feasible configurations with different values of $G(\vec{x})$.
Although these suboptimal samples do not achieve $G_{\mathrm{opt}}$, some of them still correspond to relatively high predicted growth rates.
This feature is important for practical policy analysis.
In real municipal planning, the globally optimal configuration may not be immediately achievable because of budgetary, institutional, demographic, or geographic constraints.
In such cases, feasible configurations with moderately high values of $G(\vec{x})$ can provide alternative target states that may be more realistic than the exact optimum.
The sampling distribution therefore provides information beyond the location of the optimum itself.
It offers a set of candidate indicator configurations that satisfy the imposed constraints and are predicted to improve the working-age population growth rate according to the surrogate model.
In this sense, the annealing-based approach can be used not only as an optimizer but also as a generator of feasible policy scenarios.
The present comparison is intended to validate the consistency of the QUBO formulation and to interpret the resulting solution landscape, rather than to claim a computational advantage of quantum annealing over classical methods.
For the present problem size, the exact solution can be obtained by Gurobi and can be used as a reliable reference.
However, the number of feasible configurations grows exponentially with the number of social indicators and discretization levels.
For $F$ social indicators with four levels each, the number of feasible one-hot configurations is $4^F$.
Therefore, if more indicators or finer discretization levels are included, exhaustive search becomes increasingly impractical.
In such larger settings, annealing-based methods may become useful because they can explore a large QUBO landscape and return multiple feasible candidate solutions.
Another advantage of the QUBO representation is that it retains the quadratic structure of the surrogate model.
As discussed in the previous section, the off-diagonal elements of the coefficient matrix encode pairwise interactions between social indicator levels.

Consequently, the optimized configurations are not determined only by independent contributions of individual indicators but also by second-order compatibility between different indicators.
This property is essential for interpreting combinations of social conditions that may jointly contribute to higher predicted working-age population growth.
Overall, the results in Fig.~2 show that the annealing-based optimization is consistent with the exact Gurobi solution and that the sampling distribution provides a useful set of feasible near-optimal configurations.
This solution set serves as the basis for the municipality-level single-flip analysis presented in the next section.

\subsection*{Municipality-level single-indicator sensitivity analysis}
We next consider how the predicted population growth rate changes when each municipal indicator is varied using the regression model $G(\vec{x})$. 
For each municipality, 
we compare the predicted value $G(\vec{x})$ obtained by virtually changing one indicator by one discretized level while keeping all other indicators unchanged. 
For each indicator, 
both an increase and a decrease of one level were considered whenever the resulting value remained within the valid range of the discretized levels.

In this experiment, 
we focus on three representative municipalities that exhibit nearly identical population growth rates and for which the regression predictions are reasonably close to the actual values.
These municipalities were selected because they have similar population growth rates, relatively accurate regression predictions, and nearly identical values for the major indicators identified by the QUBO analysis. 
Although the analysis is based on publicly available municipal data, the municipalities are anonymized and referred to as A, B, and C to avoid drawing attention to or making policy-related judgments about specific municipalities.
Their population growth rates are $y=-2.34$, $-2.40$, and $-2.40$, while the corresponding predicted values are $G(\vec{x})\approx -2.81$, $-1.85$, and $-2.13$, respectively. 
These three municipalities share the same levels of population density $\hat{x}_1$, average age $\hat{x}_2$, ratio of single-person households $\hat{x}_5$, and ratio of three-generation households $\hat{x}_6$. 
Among these indicators, all except $\hat{x}_5$ coincide with the levels of the optimized feature vector $\vec{x}_{\mathrm{opt}}$ obtained by solving the QUBO optimization problem. 
In particular, both population density ($\hat{x}_1=4$) and average age ($\hat{x}_2=1$), which were identified by the QUBO analysis as major contributors to population growth, already match the optimized feature vector. 
Nevertheless, these municipalities still exhibit negative population growth rates, providing an appropriate setting for evaluating the contributions of the remaining social indicators to the predicted population growth rate.

\begin{figure}[h]
    \centering
    \includegraphics[width=\linewidth]{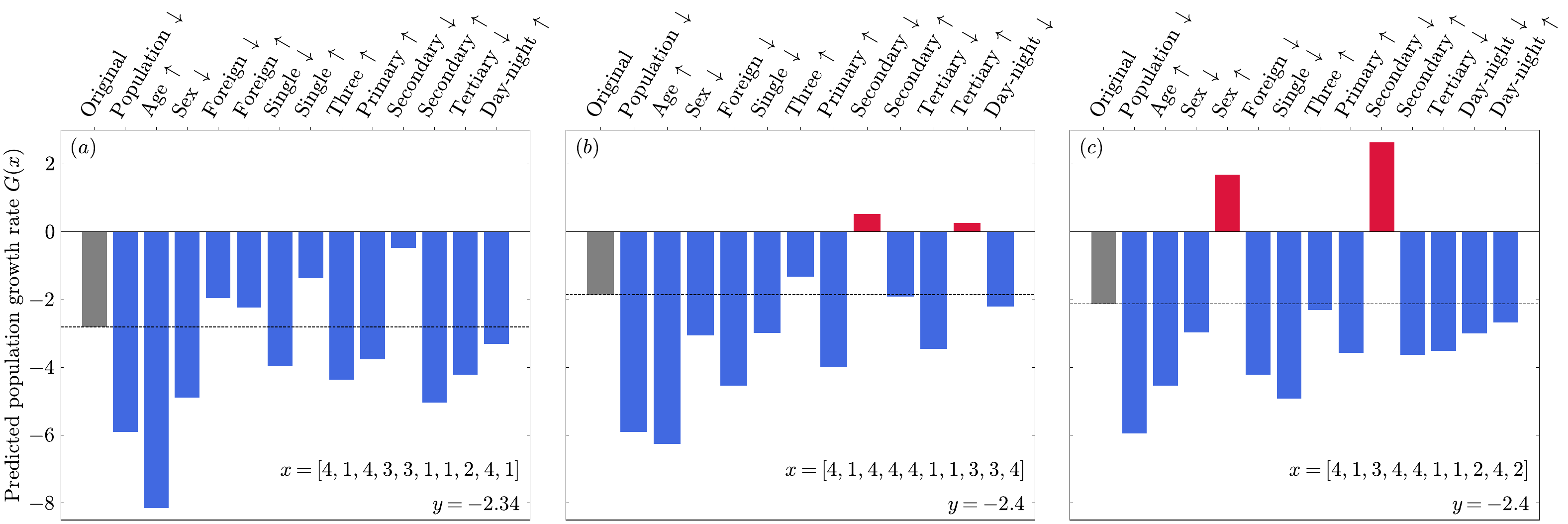}
    \caption{
        Predicted population growth rates, $G(\vec{x})$, obtained by virtually changing each municipal indicator by one discretized level for the three municipalities. The leftmost gray bar represents the prediction for the original feature vector $\vec{x}$, while the remaining bars represent the predictions obtained by changing only the corresponding indicator by one level, either decreasing (↓) or increasing (↑) it. The label 'Original' denotes the original feature vector, and the other labels correspond to the municipal indicators listed in Table~1. Blue (red) bars indicate negative (positive) predicted population growth rates. The dashed line indicates the original predicted value.
    }
    \label{fig:Example of case study}
\end{figure}

Fig.~\ref{fig:Example of case study} shows the predicted values obtained by virtually changing each indicator by one level for the three municipalities. 
A common trend is observed across all three municipalities: changing the ratio of secondary industry workers, $\hat{x}_8$, produces the largest positive change in the predicted population growth rate. 
In particular, for Municipalities B and C, the predicted value changes from negative to positive. 
These results suggest that $\hat{x}_8$ has the largest influence on the regression prediction. 
Since $\hat{x}_8=1$ in the optimized feature vector $\vec{x}_{\mathrm{opt}}$, this result is also consistent with the direction toward the optimized solution.

On the other hand, 
moving closer to $\vec{x}_{\mathrm{opt}}$ does not necessarily lead to an improvement in the predicted population growth rate. 
For example, 
in Municipalities A and B, the level of the day-night population ratio, $\hat{x}_{10}$, is lower than that of $\vec{x}_{\mathrm{opt}}$. 
However, changing this indicator toward the optimized level decreases $G(\vec{x})$. 
This observation suggests that the effect of each indicator is not determined independently, but rather depends on its combination with other indicators. 
Therefore, improving the predicted population growth rate cannot always be achieved by independently adjusting each indicator toward its optimal level. 
Instead, the combination of indicator levels should be optimized jointly. 
This behavior is consistent with the QUBO formulation, which incorporates not only the individual contributions of the indicators but also their pairwise interactions. 
These results provide empirical support for introducing pairwise interaction terms into the QUBO formulation.

Although this analysis is limited to three representative municipalities, 
it illustrates how the regression model can be used to evaluate the local effect of individual indicators and highlights the importance of considering interactions among indicators when seeking combinations of municipal characteristics associated with higher predicted population growth rates. 

\section*{Conclusion}

In this study, 
we developed an interpretable QUBO-based framework for modelling and optimizing municipal social-indicator configurations associated with working-age population growth.
Using social indicators from Japanese municipalities, we first constructed a quadratic surrogate model $G(\vec{x})$ for the working-age population growth rate.
The surrogate model showed reasonable predictive performance, with a test-set correlation coefficient of 0.84 and an average coefficient of determination of $R^{2}=0.76$.
Because the model was formulated as a quadratic function of one-hot encoded social indicators, 
its coefficients could be visualized as a matrix and interpreted in terms of both individual indicator-level contributions and pairwise interactions.

The coefficient analysis showed that the lowest quartile of average age had the largest positive diagonal contribution, 
which is qualitatively consistent with the target variable being the working-age population growth rate.
It also identified a strong positive interaction between the lowest average-age quartile and the second quartile of the primary-industry worker ratio, 
illustrating the importance of considering combinations of social conditions rather than only individual indicators.
We then optimized the QUBO representation of the surrogate model using quantum annealing, simulated annealing, and Gurobi.
Both annealing-based methods reached the same optimal value as the exact Gurobi solution in the present problem setting.
This result confirms the consistency of the QUBO formulation and shows that annealing-based searches can recover the optimal feasible configuration for the present model.

At the same time, 
the sampling distributions obtained from the annealing methods provided multiple feasible configurations with different predicted growth rates.
This is important from a practical perspective, 
because the globally optimal configuration may not be immediately achievable for actual municipalities owing to budgetary, institutional, demographic, or geographic constraints.
Thus, the proposed framework can be used not only to identify an optimum but also to generate feasible candidate configurations for model-based policy discussion.

Finally,
we performed municipality-level single-indicator flip analyses for representative municipalities.
These analyses showed that changing a single indicator can have different effects depending on the configuration of the other indicators.
In particular, moving an individual indicator closer to the globally optimized configuration did not always improve the predicted growth rate.
This result supports the use of a quadratic QUBO representation, because pairwise interactions between indicator levels are essential for interpreting why some local changes improve the predicted outcome whereas others do not.
The single-flip analysis therefore provides a local sensitivity interpretation of the fitted surrogate model and connects the global QUBO optimization results to municipality-level scenarios.

The proposed approach should be interpreted as a model-based exploratory framework rather than as a causal estimate of policy interventions.
The learned coefficients and optimized configurations represent associations captured by the fitted surrogate model, and actual policy implementation would require additional institutional, economic, and causal analyses.
Nevertheless, by combining interpretable surrogate modelling with QUBO-based optimization, this framework provides a transparent way to connect municipal social statistics, nonlinear indicator interactions, and feasible scenario generation.
Future work should examine the temporal stability of the surrogate model using data from other periods, test alternative discretization schemes, incorporate a broader set of municipal indicators, and evaluate the robustness of the optimized configurations across different modelling assumptions and regional contexts.

\section*{Acknowledgements}
This study was supported by the Cross-ministerial Strategic Innovation Promotion Program (SIP) from the Cabinet Office (No. 23836436). We would like to thank Mr. Tomonori Honda (Project Igniter \& Manager, MARUMORI-SAUNA) for providing the impetus to begin this research and for his invaluable insights into the regional issues facing Marumori, Miyagi.

\bibliography{reference}

\end{document}